\documentclass{llncs}
\usepackage[T1]{fontenc}
\usepackage{lmodern}
\usepackage[utf8]{inputenc}
\usepackage[pdftex]{graphicx}
\usepackage{array,upquote}
\usepackage{url}
\usepackage{latexsym}
\usepackage{amsmath}
\usepackage{amssymb}
\usepackage{mymacros}
\usepackage{microtype,xspace,fixltx2e}
\usepackage[usenames,svgnames,table]{xcolor}
\usepackage[bookmarks=false]{hyperref}

\hypersetup{
   colorlinks=true,
   urlcolor=blue,
   pdfborder= 1 1 1
}

\definecolor{gris}{gray}{0.5}

\usepackage{listings}
\usepackage{lstlangcoq}
\newcommand{\compcert}{CompCert}
\newcommand{\astree}{Astr\'ee}

\definecolor{mygreen}{RGB}{30,192,30}

\ifx\proof\undefined% Some styles, e.g. jfp.sty, already provide a "proof" env.
\def\proof{\trivlist \item[\hskip\labelsep {\bf Proof:}]}

\fi
\def\case#1.{\medskip\noindent {\bf Case} #1.}
\def\cas#1.{\medskip\noindent {\bf Cas} #1.}

\def\eg{e.g.,\xspace}
\def\ie{i.e.,\xspace}

\def\etal{~\emph{et al.}\xspace}

\def\opt#1{#1^?}

\def\reach#1{{\lstinline!reach!}(#1)}
\def\ofs{i} %\def\ofs{{\it ofs}}

\def\chunk{\kappa} %{\it chunk}}

\def\sig{{\it sig}}

\def\E0{{\epsilon}}

\newcommand{\id}{\ensuremath{\mathit{id}}}

\newcommand{\res}{{\it res}}

\newcommand{\op}{{\it op}}
\newcommand{\cmp}{{\it b}}

\newcommand{\valanal}{{\lstinline!value_analysis!}}

\begin{document}

\title{Formal Verification of a C Value Analysis Based on Abstract
  Interpretation
\thanks{This work was supported by Agence Nationale de la Recherche, grant
number ANR-11-INSE-003 Verasco.}
}
\author{Sandrine Blazy\inst{1} \and Vincent Laporte\inst{1} \and Andre  Maroneze \inst{1} \and David Pichardie\inst{2}}
\institute{IRISA - Universit\'e Rennes 1\and Harvard University / INRIA }
%\authorrunning{Blazy, Maroneze, Pichardie}
\maketitle

\begin{abstract}
Static analyzers based on abstract interpretation are complex
pieces of software implementing delicate algorithms.
Even if static analysis techniques are well understood, their
implementation on real languages is still error-prone. 

This paper presents a formal verification using the Coq proof
assistant: a formalization of a value analysis (based on abstract
interpretation), and a soundness proof of the value analysis. The
formalization relies on generic interfaces. The mechanized proof is
facilitated by a translation validation of a Bourdoncle fixpoint iterator. 

The work has been integrated into the CompCert verified C-compiler.
Our verified analysis directly operates over an intermediate language
of the compiler having the same expressiveness as C.  
The automatic extraction of our value analysis into OCaml yields
a program with competitive results, obtained from experiments on 
a number of benchmarks and comparisons with the Frama-C tool. 
\end{abstract}

\section{Introduction}

Over the last decade, significant progress has been made in developing
tools to support mathematical and program-analytic reasoning. Proof
assistants like ACL2, Coq, HOL, Isabelle and PVS are now successfully applied
both in mathematics (\eg a mechanized proof of the 4-colour
theorem~\cite{GonthierFC} and of the Feit-Thompson
theorem~\cite{GonthierFT}) 
and in formal verification of critical software systems (\eg the 
CompCert C-compiler~\cite{Leroy-backend}  and the verified
operating system kernel seL4~\cite{sel4}).  

Over the same time, automatic verification tools based on
model-checking, static analysis and program proof have become widely
used by the critical software industry.  
The main reason for their success is that they strengthen the
confidence we can have in critical software by providing evidence of
software correctness.
The next step is %to have more confidence in these verification tools.
to strengthen the confidence in the results of these verification tools,
and proof assistants seem to be mature and adequate for this task.
%tools involved in the production and verification of critical software
This paper presents a foundational step towards the formal
verification of a static analysis based on abstract
interpretation~\cite{CousotCousot92-2}: the formal verification using the Coq
proof assistant %~\cite{Bertot-Casteran-Coqart}
 of a value-range
analysis operating over a real-world language. 

Static analyzers based on abstract interpretation are complex pieces
of software that implement delicate symbolic algorithms and numerical
computations. Their design requires a deep understanding of the
targeted programming language. Misinterpretations of the programming
language informal semantics may lead to subtle soundness bugs that may
be hard to detect by using only testing techniques.
Implementing a value analysis raises specific issues related to
low-level numeric computations. First, the analysis must handle the machine
arithmetic that is (more or less) defined in the programming language.
Second, some computations done by the analyzer rely on this machine
arithmetic. 

Thus, a prerequisite for implementing a static analyzer operating over
a C-like language is to rely on a formal semantics of the programming language
defining precisely the expected behaviors of any program execution
(and including low-level features such as machine arithmetic).
%, involving abstract domains, transfer functions, semantics, fixpoint iterators, ???.
%The static analysis techniques are well understood but their implementation is still error-prone.
%
Such formal semantics are defined in the CompCert compiler (and it is
unusual for a compiler). More
precisely, each language of the compiler is defined by a formal
semantics (in Coq) associating observable behaviors to any
program. Observable behaviors include normal termination, divergence,
abnormal termination and undefined behaviors (such as out-of-bounds
array access).
We have chosen one language of the compiler (the main intermediate
language that has the same expressiveness as C, see
Section~\ref{sec:background}) and we have formalized a static analyzer
operating over this language. The advantage of this approach is that
our analyzer as well as the formal semantics operate exactly over the
same language. 

The main peculiarity of the CompCert C-compiler is that it is equipped
with a proof of semantic preservation~\cite{Leroy-backend}. This
proof is made possible thanks to the formal semantics of the languages
of the compiler. The proof states that any compiled program behaves
exactly as specified by the semantics of its original program. It
%relies on formal semantics of the source and target languages of the
%compiler; it 
consists of the composition of
correctness proofs for each compiler pass and thus involves reasoning
on the different intermediate languages of the compiler.
%More precisely, one of the main theorems is the following.
%
%{\itshape
%Let S be a source C program. Assume that S is free of undefined
%behaviors. Further assume that the CompCert compiler, invoked on S,
%does not report a compile-time error, but instead produces executable
%code E. Then, any observable behavior of E is one of the possible
%observable behaviors of S.
%}

%The CompCert development consists in several libraries that are
%reusable for formally verifying a static analyzer. For instance, the
%following libraries are reusable: integers, floats, memory model.
%%Other libraries need to be adapted in order to be usable for a static analyzer
%\dpnote{On peut le dire plus tard dans la papier sinon}

%We have proved the correctness of our value analysis with
%a machine-checked proof in Coq. 
%One contribution of our formalization work is to provide a correct reference 
%description of such an analysis.  We believe this is the first formal
%verification of a value analysis that operates over a real language.
%
%Moreover, 

All results presented in this paper have been mechanically verified using the
Coq proof assistant. The complete Coq development is available online at \url{http://www.irisa.fr/celtique/ext/value-analysis}.
%\footnote{The complete Coq development is available at \url{www.irisa.fr/celtique/ext/value-analysis}.}

The paper makes the following contributions.
\begin{itemize}
\item It provides the first verified value analysis for a realistic
  language such as C and hence demonstrates the usability of theorem proving in
static analysis of real programs.
\item It presents a modular design with strong interfaces aimed at
  facilitating any further extension.
\item It provides a reference description of basic techniques of
  abstract interpretation and thus gives advice on how to use the
  abstract interpretation methodology for this kind of exercice while
  maintaining a sufficiently low cost in terms of formal proof effort.
\item It compares the performances of our tool (that has been generated automatically from our
formalization and integrated into the \compcert{} compiler) with those
of two interval-based value analyzers for C.  
\end{itemize}

%The Coq development itself is designed to be as accessible as
%possible.  
The paper exposes many examples taken from the formal development.
% but we don't assume any prior knowledge in Coq and introduce any new concepts.
%To this end, we minimise the
%use of external libraries and advanced features of Coq. Where these can offer advantages, we highlight that fact and point to relevant literature. 
%
It is structured to follow the development of a C value
analysis based on abstract interpretation; from generic abstract domains (section~\ref{sec:abdom}),
 to fixpoint resolution (section~\ref{sec:fixpoint}) and numerical and
 memory abstractions (sections~\ref{sec:num-abstract} and~\ref{sec:mem-abstract}). 
%Each section explains the approach we took to formalising the relevant
%constructs in Coq and points out both, advantages as well as costs of
%formalisation, and scope for further developments.   
%
%The remainder of this paper is organized as follows. First,
%Section~\ref{sec:itv} defines the interval abstract domain.
%Then, Section~\ref{sec:syntax} briefly introduces the
%RTL intermediate language over which the value analysis is conducted. Then, \ref{sec:value-analysis}
%details our value analysis. Section~\ref{sec:proof} is devoted to its soundness proof. 
Section~\ref{sec:exper-eval} describes the experimental evaluation of our implementation.
Related work is discussed in Section~\ref{sec:related-work}, followed by concluding remarks.

%\paragraph{Notations}  $\closed x y$ denotes the closed interval of
%integers $\{ n \in \mathbb{Z} \mid x \le n \le y\}$.  
%For functions
%returning ``option'' types, $\some x$ (read: ``some $x$'') corresponds
%to success with return value $x$, and~$\None$ (read: ``none'')
%corresponds to failure.  
%In grammars, $\seq a$ denotes 0, 1 or several
%occurrences of syntactic category $a$, and $\opt a$ denotes an
%optional occurrence of syntactic category~$a$. 

\section{Background}\label{sec:background}

This section starts with a short introduction to the Coq proof
assistant. It is followed by a brief presentation of the CompCert
architecture and memory model. The language our analyzer operates over
is described at the end of this section.

\subsection{Short Introduction to Coq}
Coq is an interactive theorem prover. It consists in a strongly typed
specification language and a language for conducting machine-checked
proofs interactively.  The Coq specification language is a functional
programming language as well as a language for inductively defining
mathematical properties, for which it has a dedicated type
(\lstinline!Prop!).  Induction principles are automatically generated
by Coq from inductive definitions, thus inductive reasoning is very
convenient.  Data structures may consist of properties together with
dependent types. Coq's type system includes type classes.  Coq
specifications are usually defined in a modular way (\eg using record
types and functors, that are functions operating over structured data
such as records). 
%In a Coq specification, some definitions may be left
%abstract (unspecified).  
%Thus, a Coq development may be parameterized
%by such abstract definitions.  
The user is in charge to interactively
build proofs in the system but those proofs are automatically
machine-checked by the Coq kernel. OCaml programs can be automatically
generated by Coq from Coq specifications. This process is called extraction.

\subsection{The CompCert Memory Model}\label{sec:compc-memory-model}

There are 11 languages in the CompCert compiler, including 9
intermediate languages. These languages feature both low-level aspects
such as pointers, pointer arithmetic and nested objects, and
high-level aspects such as separation and freshness guarantees.
A memory model~\cite{Leroy-Appel-Blazy-Stewart-memory-v2} is shared by the semantics of all these languages.
%It is defined in~\cite{Leroy-Appel-Blazy-Stewart-memory-v2}. %Leroy-Blazy-memory-model,
Memory states (of type \lstinline!mem!) are collections of blocks, each block being an array
of abstract bytes. A block represents a C variable or an invocation of \lstinline!malloc!.
Pointers are represented by pairs \lstinline!(b,i)! of a block identifier and
a byte offset  \lstinline!i! within this block.
Pointer arithmetic modifies the offset part of a pointer value,
keeping its block identifier part unchanged.
%In CompCert, the semantics for C 
%\cite{Blazy-Leroy-Clight-09} associates a different block to every
%global variable of the program, to every addressable local variable of every
%active invocation of a function of the program, and to every
%invocation of {\tt malloc}.  For local variables, fresh blocks are
%allocated at function entry point and deallocated when the function
%returns.

%$$ (b, \ofs) + n \defequal (b, \ofs + n) $$
%As a consequence, blocks are separated by construction: from a pointer
%to block $b$, no amount of pointer arithmetic can create a pointer to
%block $b' \not= b$; pointer arithmetic can only create other pointers
%within block $b$, or illegal pointers outside $b$'s bounds.

Values stored in memory are the disjoint union of 32-bit
integers (written as  \lstinline!vint(i)!), 64-bit floating-point numbers,
locations (written as \lstinline!vptr(b,i)!), and the special value \lstinline!undef! representing the
contents of uninitialized memory. Pointer values \lstinline!vptr(b,i)! are
composed of a block identifier \lstinline!b! and an integer byte offset \lstinline!i! within this block. 
%\begin{syntaxleft}
%v & ::=  & {\tt vint}(i) \mid {\tt vfloat}(f) \mid {\tt vptr}(b,\ofs) \mid {\tt undef}
%\end{syntaxleft}
%
Memory chunks appear in memory operations \lstinline!load! and
\lstinline!store!, to describe concisely the size, type and signedness of the value being stored.
\begin{syntaxleft}
\syntaxclass{Values:}
v & ::=  & {\tt vint}(i) \mid {\tt vfloat}(f) \mid {\tt vptr}(b,\ofs) \\
& \mid & {\tt undef}
\syntaxclass{Mem. chunks:}
\chunk & ::= & {\tt Mint8signed} \mid {\tt Mint8unsigned} & 8-bit integers \\
     & \mid& {\tt Mint16signed} \mid {\tt Mint16unsigned} & 16-bit integers \\
     & \mid& {\tt Mint32} & 32-bit integers or pointers \\
     & \mid& {\tt Mfloat32} & 32-bit floats \\
     & \mid& {\tt Mfloat64} & 64-bit floats
\end{syntaxleft}

In CompCert, a 32-bit integer (type \lstinline!int!) is defined as 
a Coq arbitrary-precision integer (type \lstinline!Z!) plus a property called
\lstinline!intrange! that it
is in the range $0$ to $2^{32}$ (excluded). 
%There are signed and
%unsigned integers in C programs. Due to (implicit and explicit) casts
%(conversions) between C variables, 
The function
\lstinline!signed! (resp.~\lstinline!unsigned!) gives an interpretation of machine 
integers as a signed (resp.~unsigned) integer. The properties 
\lstinline{signed_range} and \lstinline{unsigned_range} are examples
of useful properties for machine integers.
% In the sequel of this
%paper, we use the notation
%{\tt Int.XX} to denote the function called {\tt XX} of the
%module called {\tt Int}.
%\sbnote{same notation for accessing the filed of a record}

%\begin{figure}
\begin{lstlisting}
Definition max_unsigned : Z := $2^{32}$ - 1.  
Definition max_signed : Z := $2^{31}$  - 1.
Definition min_signed : Z := - $2^{31}$.
Record int := {   intval: Z;
                  intrange: 0 $\leq$ intval < $2^{32}$ }.
Definition unsigned (n: int) : Z := intval n.
Definition signed (n: int) : Z := if unsigned(n) < $2^{31}$ then unsigned(n) 
                                  else unsigned(n) - $2^{32}$.
Theorem signed_range: forall i, min_signed $\leq$ signed(i) $\leq$ max_signed.
  Proof.  (* Proof commands omitted here *)  Qed.
Theorem unsigned_range: forall i, 0 $\leq$ unsigned(i) $\leq$ max_unsigned.
  Proof.  (* Proof commands omitted here *)  Qed.
\end{lstlisting}
%\caption{\label{fig:int}}
%\end{figure}

\subsection{The CFG Intermediate Language}

The main intermediate language of the CompCert compiler is called
Cminor, a low-level imperative language structured like C into
expressions, statements and functions. Historically, Cminor was the target
language of the compiler front-end. There are four main differences
with C~\cite{Leroy-backend}. First, arithmetic operators are not
overloaded. Second, address computations are explicit, as well as memory access
(using load and store operations). Third, control
structures are if-statements, infinite loops, nested blocks
plus associated exits and early returns. Last, local variables can only
hold scalar values and they do not reside in memory, making it
impossible to take a pointer to a local variable like the C operator
{\tt \&} does. Instead, each Cminor function declares the size of a
stack-allocated block, allocated in memory at function entry and
automatically freed at function return. 
The expression \lstinline!addrstack(n)! returns a pointer within that
block at constant offset \lstinline!n!.
%\sbnote{probablement a racourcir: explicit stack allocation of certain
%local variables}

Cminor was designed to be the privileged language for integrating within
CompCert other tools operating over C and other compiler front-ends. For
instance, two front-end compilers from functional languages to Cminor
have been connected to CompCert using Cminor, and a separation logic
has been defined for Cminor~\cite{Appel-Blazy-07}. 
The Concurrent Cminor language extends Cminor with concurrent features
and lies at the heart of the Verified Software Toolchain project~\cite{VST}.

As control-flow is still complex in Cminor (due to the presence of
nested blocks and exits), we have first designed a new
intermediate language called CFG that is adapted for static analysis:
1) its expressions are Cminor expressions (\ie side-effect free
C expressions), 2) its programs are represented by their control flow graphs, with
explicit program points and 3) the control flow is restricted to
simple unconditional and conditional jumps.
%CFG represents programs by their control-flow graph. 
The CFG syntax is defined in Figure~\ref{fig:syntax}. Floating-point operators
are omitted in the figure, as our analysis does not compute any information about floats.
Statements include assignment to local variables, memory stores,
if-statements and function calls. 
Expressions include reading local variables, constants and arithmetic
operations, reading store locations, and conditional expressions. 
As in the memory model, loads and stores are parameterized by a memory
chunk $\chunk$.

\begin{figure}
\begin{syntaxleft}
\syntaxclass{Constants:}
 c & ::= & n \alt f & integer and floating-point constants\\
   & \alt & {\tt addrsymbol}(\id,n) &  address of a symbol plus an offset\\
   &\alt & {\tt addrstack}(n) & stack pointer plus a given offset

\syntaxclass{Expressions:}
a & ::=  & \id & variable identifier\\
  & \alt & c & constant\\
  & \alt & \op_1 ~ a & unary arithmetic operation\\
  & \alt & a_1 ~\op_2 ~ a_2 \ & binary arithmetic operation\\
  & \alt & a_1 ? ~ a_2 : a_3 \ & conditional expression\\
  & \alt & {\tt load} (\chunk,a) & memory load 

\syntaxclass{Unary op.:}
\op_1 & ::= & {\tt cast8unsigned} & 8-bit zero extension \\
 & \alt & {\tt cast8signed} & 8-bit sign extension \\
          & \alt & {\tt cast16unsigned} & 16-bit zero extension \\
          & \alt & {\tt cast16signed} & 16-bit sign extension \\
          & \alt & {\tt boolval} & 0 if null, 1 if non-null \\
          & \alt & {\tt negint} & integer opposite \\
          & \alt & {\tt notbool} & boolean negation \\
          & \alt & {\tt notint} & bitwise complement 

\syntaxclass{Binary op.:}
\op_2 & ::=  & \hbox{{\tt +}} \alt \hbox{{\tt -}} \alt \hbox{{\tt *}} \alt \hbox{{\tt /}} \alt \hbox{{\tt \%}}
               & arithmetic integer operators \\ 
      & \alt & \hbox{{\tt <\/<}} \alt \hbox{{\tt >\/>}} \alt 
               \hbox{{\tt \&}} \alt \hbox{{\tt {\char124}}} \alt \hbox{{\tt {\char94}}}
               & bitwise operators\\
          & \alt & \hbox{{\tt /}}_u \alt \hbox{{\tt \%}}_u \alt \hbox{{\tt >\/>}}_u & unsigned operators \\
          & \alt & {\tt cmp}(\cmp)  & integer signed comparisons\\
          & \alt & {\tt cmpu}(\cmp) & integer unsigned comparisons

\syntaxclass{Comparisons:}
\cmp &  ::=  & \hbox{{\tt <}} \alt \hbox{{\tt <=}} \alt \hbox{{\tt >}} \alt \hbox{{\tt >=}} 
               \alt \hbox{{\tt ==}} \alt \hbox{{\tt !=}}   & relational operators 

\syntaxclass{Statements:}
i & ::=  & {\tt skip} (l)& no operation (go to l)\\
  & \alt & {\tt assign} (\id,a, l) & assignment\\
  & \alt & {\tt store} (\chunk, a,a, l) & memory store \\ 
  & \alt & {\tt if} (e, l_{true}, l_{false})  & if statement\\
  & \alt & {\tt call} (\sig,\opt{\id}, a,a*,l) & function call \\
  & \alt & {\tt return} \opt{(a)} & function return

\end{syntaxleft}
\caption{Abstract syntax of CFG}
\label{fig:syntax}
\end{figure}

%          & \alt & {\tt absf} \alt {\tt singleoffloat} & float absolute value and truncation \\
%          & \alt & {\tt intoffloat}  \alt {\tt intuoffloat} & signed and unsigned integer to float\\
%          & \alt & {\tt floatofint} \alt {\tt floatofintu} & float to signed and unsigned integer 
%
%           & \alt & {\tt cmpf}(\cmp) & float comparison

\begin{figure}
  \centering
\includegraphics[width=.8\textwidth]{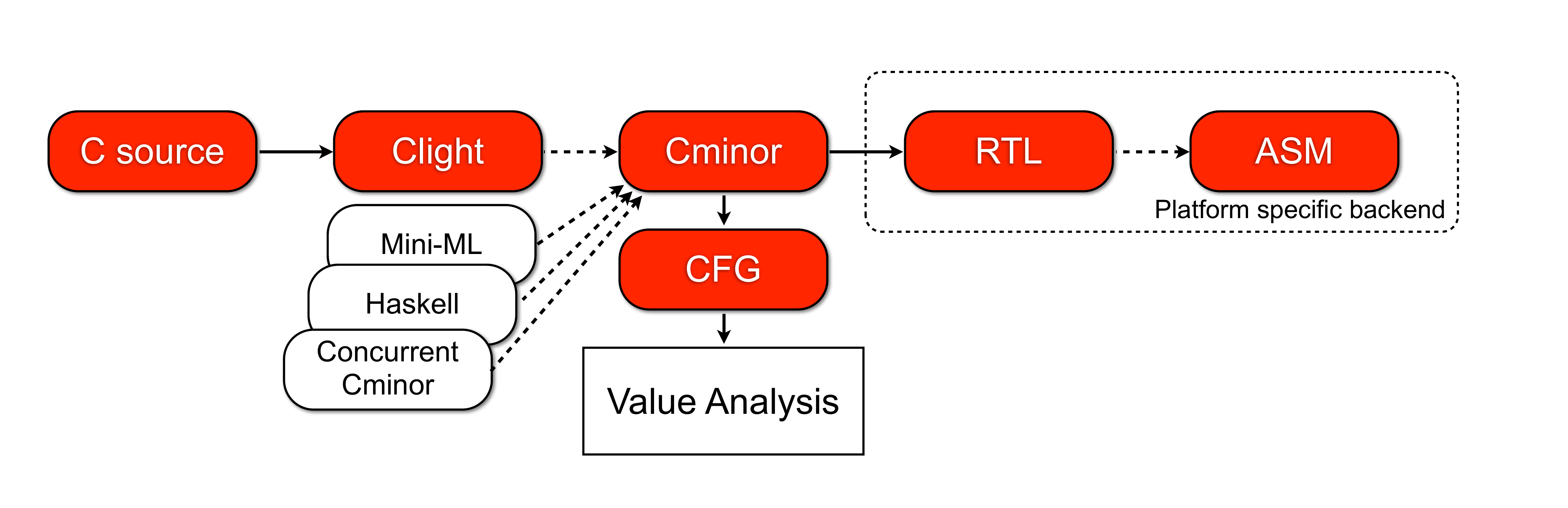}  
\caption{Integration of the value analysis in the CompCert toolchain}
 \label{fig:chain}
\end{figure}

The CFG language is integrated into the CompCert compiler, as shown in Figure~\ref{fig:chain}.
There is a translation from Cminor to CFG and a theorem stating that
any terminating or diverging execution of a CFG program is also a
terminating or diverging execution of the original Cminor program.
Thus, instead of analyzing Cminor programs, we can analyze CFG programs
and use this theorem to propagate the results of the CFG analysis on Cminor programs.
%\vlnote{A Cminor program… the corresponding CFG program}
For instance, in order to show that Cminor is memory safe, we only
need to show that CFG is memory safe. %The integration of this new language in the
%CompCert toolchain is not the topic of this paper. It is due to Jacques-Henry Jourdan.\dpnote{remercier JH comme cela ?}

%Our static analysis operates on the CFG intermediate language, that has the same expressiveness as C.

%Cast operations used by arithmetic conversions rules in C expressions are made explicit at the CFG level.
%\sbnote{Expliquer les addr dans les constantes ?}
%Among expressions are several comparisons.
%In C, there is for instance only one greater-than operator; it represents different operators that are made explicit in the CFG language: 
%they are specific to either signed integers, or to unsigned integers or to floats.

For the purpose of the experiments that we conduct in
Section~\ref{sec:exper-eval}, we use an inlining pass recently added
to the CompCert compiler. It was implemented and
proved correct by X.Leroy for another language
of the compiler, RTL, that is similar to CFG except
that it only handles flat expressions. Since our analysis operates on CFG, we have adapted this 
inlining pass to CFG. Adapting
the soundness proof of this transformation to CFG has been left for future work.

The concrete semantics of CFG is defined in small-step style as a
transition relation between execution states.  
An execution state is a tuple called $\sigma$. Among the components of
$\sigma$ are % made of a call stack, the current function, 
the current program point (\ie a node in the control-flow graph), 
the memory state (type \lstinline!mem!) and the environment
(type \lstinline!env!) mapping program variables to values.  
We use $\sigma.E$ to denote the environment of a state $\sigma$,
and ${\tt dom}(\sigma.E)$ to denote its domain (\ie the set of its
variables). We use 
%$\trexec{P}$ to denote sets of execution
%traces of a program $P$, and
$\reach{P}$ to denote the set of states belonging to the
execution trace of $P$.

Our value analysis (called {\tt \valanal{}}) computes
%at each program point of a program, an over-approximation
%of the domain size of each variable.
%
%The value analysis computes 
for each program point the estimated values of the program variables.
When the value of a variable is an integer \lstinline!i!  or a pointer
value of offset \lstinline!i!, the estimate provides two numerical
ranges \lstinline!signed_range! and
\lstinline!unsigned_range!. The first one over-approximates the signed interpretation of
\lstinline!i! and the other range over-approximates its unsigned
interpretation. We note \lstinline!ints_in_range (signed_range,unsigned_range) i! this fact.
Thus, given a program $P$,  $\valanal{}(P)$ yields a map
%\vlnote{Donner un nom à la map?}
such that for each node $l$ in its control flow graph and
 each variable $v$,
$\valanal{}(P)[l,v]$ is a pair of sound ranges for $v$.
%an interval $[a,b]$ %(of type $\mathtt{itv}$)
%that represents a conservative range of the possible values of $v$ at $l$.
 %
The following theorem states the soundness of the value analysis: for
every program state that may be reached during the execution of a program,
any program point and variable, every variable valuation computed
by the analysis is a correct estimation of the exact value
given by the concrete semantics.

\begin{theorem}[Soundness of the value analysis]
Let $P$ be a program, $\sigma\in\reach{P}$ and
$\res=\valanal{} (P)$ be the result of the value
analysis. Then, for each program point $l$, for
each local variable
$v\in{\tt dom}(\sigma.E)$
that contains an integer $i$ {\textup(}\ie
$  \sigma.E(v) = \mathtt{vint}(i) \vee \exists b, \sigma.E(v) = \mathtt{vptr}(b,i) ${\textup)},
the property $(\mathtt{ints\_in\_range}~\res[l,v] ~ i)$ holds. 
\end{theorem} 
%\sbnote{a revoir ?}\dpnote{Dans la suite, on affine un peu en parlant de sign/unsign}

\subsection{Overview of a Modular Value Analysis}

Our value analysis is designed in a modular way: a generic fixpoint iterator
operates over generic abstract domains (see Section~\ref{sec:abdom}). The iterator is based on the
state-of-the-art Bourdoncle~\cite{bourdoncle93} algorithm that provides both efficiency
and precision (see Section~\ref{sec:fixpoint}).

\begin{figure}
  \centering
\includegraphics[width=.8\textwidth]{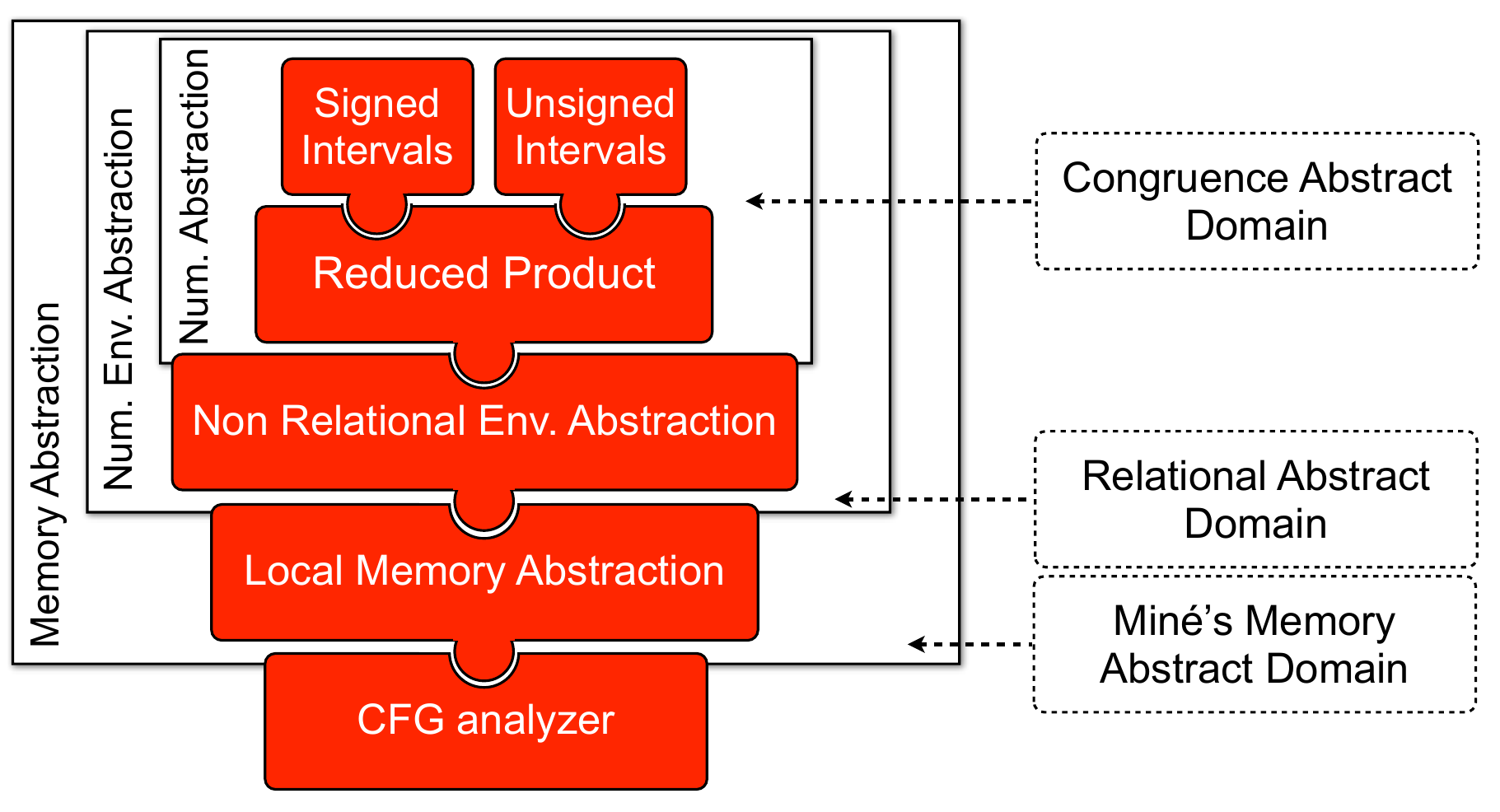}  
\caption{Design of abstract domains: a three-layer view \label{fig:couches}}
\end{figure}

The modular design of the
abstract domains is inspired from the design of the Astrée analyzer. It
consists in three layers that are showed in Figure~\ref{fig:couches}. 
%and detailed in the next section.
The simplest domains are numerical abstract domains made of intervals
of machine integers. These domains are not aware of the C memory model.

In a C program, a same piece of data can be used both in signed and
unsigned operations, and the results of these operations differ from
one interpretation to the other.  Thus, we have two numerical abstract
domains, one for each interpretation.  Our analysis computes the
reduced product of the two domains in order to make a continuous
fruitful information exchange between these two domains (see Section~\ref{sec:num-abstract}).

Then, we build abstract domains representing numerical environments.
We provide a non-relational abstraction that is parameterized by a numerical abstract domain.
The last layer is the abstract domain representing memory. It is
parameterized by the previous layer and links
the abstract interpreter with the numerical abstract domains (see Section~\ref{sec:mem-abstract}). 

This modular design is targeted to connect at each layer other
abstract domains.  They are represented in dotted lines in
Figure~\ref{fig:couches}.  For example, several abstract memory models
can be used instead of the current one while maintaining the same
interfaces with the rest of the formal development.  The ultimate goal
is to enhance our current abstract interpreter in order to connect it to a memory
domain \emph{à la} Miné~\cite{mine-LCTES06}.  The current interfaces are
also compatible with any relational numerical abstract domain.  At
the top, more basic numerical abstractions as congruence could be
added and plugged into our reduced product.

\section{Abstract Domain Library}\label{sec:abdom}

This section describes the library we have designed to represent our
abstract domains. First, it defines generic abstract domains. Then, it
details the interval abstract domain. Last, it explains how to combine
abstract domains.

\subsection{Abstract Domain Interface}

Abstract interpretation provides various
frameworks~\cite{CousotCousot92-2} for the design of abstract %,Cousot06ASIAN
semantics. The most well-known framework is based on Galois
connections but some relaxed frameworks % ~\cite{CousotCousot79-1}
exist. They are
generally used when some useful abstraction does not fulfill standard
properties (\eg polyhedral abstract
domains~\cite{CousotHalbwachs78-POPL} do not form a complete
lattice). In our context, a relaxed framework is required because of the
associated lightweight proof effort. 

Since our main goal is to provide a formal proof of soundness
for the result of an analysis, some additional properties such as best
approximation or completeness do not require a machine checked proof. 
In some previous work of the last author, a framework has been defined
for the purpose of machine checked proofs~\cite{Pich:these}. In this
paper, we push further this initiative and provide a more minimalist
framework.
The signature of abstract domains is of the following form.
\footnote{In this paper, for the sake of simplicity, we only use records to
  structure our formalization. However, in our development, we also
  use more advanced Coq features such as type classes.}
%Lattices are a basic data structure in abstract interpretation.
%The standard signature for minimal lattices is of the following form.
\begin{lstlisting}
Notation ℘(A) := (A -> Prop).       (* identify sets and predicates *)
Notation x ∈ P := (P x).
Notation P1 ⊆ P2 := (incl P1 P2).  (* property inclusion *)

Record adom (A:Type) (B:Type) : Type := {
  le: A -> A -> bool;                  (* partial order test        *)
  top: A;                              (* greatest element          *)
  join: A -> A -> A;                   (* least upper bound         *)
  widen: A -> A -> A;                  (* widening operator         *)
  gamma: A -> ℘(B);                    (* concretization function   *)
  gamma_monotone:                      (* monotonicity of gamma     *)
   forall a1 a2, le a1 a2 = true -> (gamma a1) ⊆ (gamma a2);
  top_sound:                           (* top over-approximates any *)
     ∀ x, x ∈ (gamma top);             (*   concrete property       *)
  join_sound: ∀ x y:A,                 (* join over-approximates    *)
    (gamma x) ∪ (gamma y) ⊆ gamma (join x y);  (* concrete union  *)
}.
\end{lstlisting}

Here, \lstinline!A! is the type of abstract values, \lstinline!B! is the type
of concrete values, and the type of the abstract domain is 
\lstinline!(adom A B)!. This type is a record with various operators (described on the right part)
 and properties about them. 
%\lstinline!A! is equipped with a comparison operator 
%called \lstinline!le!. \lstinline!top! is a constant of type \lstinline!A!. Three other 
%operations are provided (\lstinline!join!, \lstinline!widen! and \lstinline!gamma!), 
This record contains only three properties: the monotonicity
of the \lstinline!gamma! operator, the soundness of the \lstinline!top!
element and the soundness of the least upper bound operator
\lstinline!join!. We do not provide formal proof relating the abstract
order with \lstinline!top! or \lstinline!join!. Indeed any
\emph{weak-join} %~\cite{SankaranarayananCSM06}
 will be suitable here. The lack of properties
about the widening operator is particularly surprising at first
sight. In fact, as we will explain in Section~\ref{sec:fixpoint}, the
widening operator is used only during fixpoint iteration and this
step is validated \emph{a posteriori}. Thus, only the result
of this iteration step is verified and we don't need a widening
operator for that purpose. 

%These two properties are the minimal properties we need for proving
%the soundness of our analysis. Thus, no property related to \lstinline!leb!,
%\lstinline!join! or \lstinline!widen! is required.
% \sbnote{Parmi les choses que nous ne prouvons pas.
%   - leb est un ordre partiel
%   - join est une least upper bound pour cet ordre partiel
%   - widen est un widening
%   - le treillis est complet (souvent pas prouvable dans la logique constructive de Coq d'ailleurs)
%   - gamma est une moitié de connexion de Galois (un meet-morphisme). La fonction alpha n'est généralement pas programmable en Coq.}

The \lstinline,gamma, operator of \emph{every} abstract domain will be noted~\(\gamma\).
The type class mechanism enables Coq to infer which domain it refers to.

%Thanks to type classes, generic abstract domains can be reused to
%define similar abstract domains including a
%bottom element and sharing the same common fields (\eg \lstinline!gamma!).

%We instantiate this abstract domain type in different ways. In the
%overall our current verified value
% analysis manipulates X abstract domains.\dpnote{A completer plus tard}

\subsection{Example of Abstract Domain: Intervals}

Our value analysis operates over compositions of abstract domains.
The most basic abstract domain is the domain of intervals.
%
%Last, we define instances of abstract domains where the
%types {\tt A} and {\tt B} are defined: {\tt B} as the type {\tt int} of
%machine integers.
%The first instance is required to track types during our analysis in
%order disambiguate several operations on pointer and integer
%values. In this instance, {\tt A= VInt | top}. The tracking
%consists in a very simple must be {\tt VInt} analysis. This instance
%is detailled in the appendix in Figure~\ref{fig:lattice-int}.
%
%The second instance defines the abstract domain of intervals made of
%machine integers (modulo $2^{32}$).
%It is defined.
%It is detailed in Figure~\ref{fig:lattice-itv} in the following subsection. 
%
%\subsection{Abstract domain of  intervals made of machine integers}
%
Figure~\ref{fig:lattice-itv} defines the abstract domain of intervals
made of machine integers, that are interpreted as signed
integers. This instance is called \lstinline!signed_itv_adom!. The 
definitions are standard and only some of them are detailed in the
figure. An interval represents the range of the signed interpretation
of a machine integer. 
Thus, \lstinline,top, is defined as the largest interval with bounds 
\lstinline!min_signed! and \lstinline!max_signed!. 
The concretization is defined as follows. A machine integer \lstinline!n! belongs to
the concretization of an interval \lstinline!itv! iff \lstinline!signed(n)! belongs to \lstinline!itv!.
The proof of the lemma \lstinline!top_sound! follows from the 
  \lstinline!signed_range! theorem given in Section~\ref{sec:compc-memory-model}. 
%The proof of the lemma \lstinline!gamma_monotone!
%results from .

\begin{figure}%[t]
\begin{lstlisting}
Record itv := {min: Z; max: Z}.

Definition signed_itv_adom : adom itv int := {
   le := (λ itv1 itv2, ...);          (* definition omitted here *)
   top := { min:= min_signed; max:= max_signed};
   join := (λ itv1 itv2, ...);        (* definition omitted here *)
   widen := (λ itv1 itv2, ...);       (* definition omitted here *)
   gamma := (λ itv n, itv.min $\le$ signed(n) $\le$ itv.max);
   top_sound := (...);                (* proof term omitted here *) 
   gamma_monotone := (...);           (* proof term omitted here *) 
   join_sound := (...);               (* proof term omitted here *) 
}.
\end{lstlisting}
\caption{An instance called \lstinline!signed_itv_adom!: the domain of
  intervals (made of signed machine integers) with a concretization to \lstinline!$\mathcal{P}$(int)!.}
\label{fig:lattice-itv}
\end{figure}

% In an interval, integers are interpreted as signed integers. \sbnote{a revoir}
% \sbnote{explain: we lose precision when going from signed to unsigned
%   and vice versa}

We also define a variant of this domain with a concretization 
using an unsigned interpretation of machine integers:
%\begin{center}
           \lstinline!(λ itv n, itv.min$\le$unsigned(n)$\le$itv.max)!.
%\end{center}
As explained in Section~\ref{sec:num-abstract}, combining both domains recovers some precision that may be lost when
using only one of them. 
% As explained in~\cite{GustafssonEL05}, our
%\sbnote{un peu rapide ?}
%analysis should compute two intervals for each variable (one for a
%signed interpretation and another interval for an unsigned one). Here,
%we consider only signed interpretations.

The \lstinline!itv! record type provides only lower and
upper bounds of type \lstinline!Z!.  Using the expressiveness of the
Coq type system, we could choose to add an extra field requiring a
proof that \lstinline,min $\le$ max, holds. While elegant
at first sight, this would be rather heavyweight in practice, since
we must provide such a proof each time we build a new interval. For
the kind of proofs we perform, if such a property was
required, we would generally have an hypothesis of the form
\lstinline,i ∈ (γ itv), in our context and it would trivially imply that
\lstinline,itv.min $\le$ itv.max, holds.

%Note also that the analyser performs computation with integer of arbitrary precision (type \lstinline!Z!).
%This data structure avoids 

%% \paragraph{Finite sets}
%% 
%% TODO...
%% 
%% Pour l'instant nous n'utilisons pas ce domaine. Il est utile pour le modele memoire V1.
%% 
%% Nous pourrions parler de son efficacite. Mais il faudrait s'en servir.\dpnote{A decider au dernier moment}
%% 

\subsection{Abstract Domain Functors}

Our library provides several functors that build complex abstract
domains from simpler ones.
%
%Thanks to type classes, generic abstract domains can be reused to
%define similar abstract domains including a
%bottom element and sharing the same common fields (\eg \lstinline!gamma!).
%
%We reuse this abstract domain in different ways.

\paragraph{Direct Product}
A first example is the product \lstinline!(adom (A*A') B)! of two abstract
domains \lstinline!(adom A B)! and \lstinline!(adom A' B)!, where the
concretization of a pair \lstinline!(a,a'):A*A'! is the intersection
\lstinline!(γ a) ∩ (γ a')!.
%\begin{lstlisting}
%    Prod.make (A A' B: Type): adom A B -> adom A' B -> adom (A*A') B
%\end{lstlisting}
%each operation
%is defined from the operations of both initial abstract domains, and
%similarly both properties are proved from the corresponding properties
%of both intial abstract domains.

\paragraph{Lifting a Bottom Element}
A bottom element is not mandatory in our definition of abstract domains
because some sub-domains do not necessarily contain one. For instance, the
domain of intervals does not contain such a least element.
Still in our development, the bottom element plays a specific and important role
since we use it for reduction. We hence introduce a polymorphic type 
\lstinline,A+⊥, that lifts a type \lstinline!A! with an extra bottom
element called \lstinline!Bot!.
We then define a simple functor \lstinline!lift_bot! that lifts any
domain \lstinline!(adom A B)! on a type \lstinline!A! to a domain on
\lstinline!A+⊥!. 
In this new domain, the concretization function extends
the concretization of the input domain and \lstinline!γ Bot = $\emptyset$!. 
\begin{lstlisting}
Definition botlift (A:Type): Type := Bot | NotBot (x:A).
Notation A+⊥ := (botlift A).
Definition lift_bot (A B: Type): adom A B -> adom (A+⊥) B :=
    (* definition omitted here *)
\end{lstlisting} 
%
%\begin{lstlisting}
%\end{lstlisting}

\paragraph{Finite Reduced Map}
Lifted domains are used for instance as input for an important functor
of finite maps. CompCert uses intensively the \lstinline!TREE!
interface.  Given an implementation \lstinline!T! of the interface
\lstinline!TREE! and a type \lstinline!A!, an element of type
\lstinline!(TREE.t A)! represents a partial map from keys (of type
\lstinline!T.elt!) to values of type \lstinline!A!. The interface is
implemented for several kinds of keys in the
CompCert libraries. In our development, we use it to map
variables to abstract values, but also program points to abstract
environments. The functor implements the following type.
\begin{lstlisting}
AbTree.make(T:TREE)(A B:Type): adom A B -> adom (T.t A)+⊥ (T.elt -> B)
\end{lstlisting}
An element in \lstinline!(T.t A)+⊥! is turned into a function of type
\lstinline!T.elt -> A+⊥! via the function \lstinline!get! that satisfies the
following equations.
\begin{lstlisting}
         get(Bot)(k)      = Bot
         get(NotBot m)(k) = top          (* if m[k] is undefined *)
         get(NotBot m)(k) = NotBot m[k]  (* otherwise *)
\end{lstlisting}
%When feeding such a map with elements of an abstract domain, we expect two
%properties. First, if we put \lstinline! 
As a consequence,
%\begin{itemize}
%\item  
the \lstinline!top! element is represented in a lazy
way: a key is associated to it as soon as it is not bound in the
partial map. Furthermore,
%\item 
the map is reduced w.r.t. the bottom element of the input domain:
as soon as we try to bind a key to the bottom element, the whole map is shrunk to \lstinline!Bot!.
%\end{itemize}
This situation is interesting for dead code elimination and
more generally for the whole precision of an analysis.

\section{Fixpoint Resolution}\label{sec:fixpoint}

From a proof point of view, the main lesson learned from the CompCert
experiment is the following.
%\dpnote{Cette partie peut etre deplacer ailleurs si on veut racourcir}
When formally verifying a complex piece of software relying on
sophisticated data structures and delicate algorithms, it is not realistic to write the whole
software using exclusively the specification language of the proof assistant.
A more pragmatic approach to formal verification % is called translation validation in the
%literature~\cite{pnueli,Necula}. It 
consists in reusing an existing
implementation %(that does not depend on the proof assistant) 
in order to
separately verify its results. This approach is not optimal, but it is worthwhile
when the algorithm is a sophisticated piece of code and when the formal verification of each
of the results is much easier than the formal verification of the
algorithm itself.
% If you’re worried about completeness or optimality, this general
% approach is unlikely to work, but here, the aim is to get
% pragmatically valuable intervals, and validation is clearly the right
% way to go. 

The CompCert compiler combines both approaches in order to facilitate
the proofs. Most of the compiler
passes are written and proved in Coq. A few compiler passes (\eg the register 
allocation~\cite{Rideau-Leroy-regalloc}) % and some optimizations
%such as software pipelining~\cite{Tristan-Leroy-softpipe}) 
are not
written directly in Coq, but formally verified in Coq by a translation validation
approach. 
Our value analysis also combines both approaches.
% and we explain why some parts of our static analyzer use translation validation.
%We have implemented efficiently in OCaml some algorithms and we 
We have formally verified a checker
that validates \emph{a posteriori} the untrusted results of 
%an OCaml program. 
%More precisely, we have validated  \emph{a posteriori} 
%two algorithms: 1) the efficient Bourdoncle algorithm for building loop
%scopes, 2) the 
a fixpoint engine written in OCaml, that finds fixpoints using widening and
narrowing operators. 

%A difficulty concerns fixpoint resolution. 
As many data flow
analyses, our value analysis can be turned into the fixpoint
resolution of an equation system on a lattice. CompCert already
provides a classical Kildall iteration framework to iteratively find the
least fixpoint of an equation system. But using such a framework is
impossible here for two reasons. First, the lattice of bounded
intervals contains very long ascending chains that make standard Kleene
iterations too slow. Second, the non-monotonic nature of
widening and narrowing makes fixpoint iteration sensible to the iteration
order of each equation.
%\dpnote{si on a de la place on peut donner les signature de Kildall pour comparer} 

We have then designed a new fixpoint resolution framework that relies
on the general iteration techniques defined by
Bourdoncle~\cite{bourdoncle93}. First, Bourdoncle provides a strategy
computation algorithm based on Tarjan's algorithm to compute strongly
connected subcomponents of a directed graph and find loop headers for
widening positioning. This algorithm also orders each strongly
connected subcomponent in order to obtain an iteration strategy that
iterates inner loops until stabilization before iterating outer loops.
Bourdoncle then provides an efficient fixpoint iteration algorithm
that iterates along the previous strategy and requires a minimum
number of abstract order tests to detect convergence.

This algorithm relies on advanced reasoning in graph theory and
formally verifying it would be highly challenging. 
This frontal approach would also certainly be too rigid because
widening iteration requires several heuristic adjustments to reach a
satisfactory precision in practice (loop unrolling, delayed widenings, 
decreasing iterations). We have therefore opted for a more flexible
verification strategy: as Bourdoncle strategies, fixpoints are computed
by an external tool (represented by the function called
  \lstinline!get_extern_fixpoint!) and we only formally verify a fixpoint
checker (called \lstinline!check_fxp!).  

%\vlnote{J’ai mis «~analyze~» plutôt que solve-pfp comme nom de
%  programme.}
%\sbnote{Bof, il n'est question que d'analyseurs dans ce papier !}
Our fixpoint analyzer is defined below, given an abstract
domain~\lstinline!ab!, a program~\lstinline,P, and its entry point
\lstinline,entry,, the transfer functions~\lstinline!transfer! and
initial abstract values~\lstinline!init!. 
\begin{lstlisting}
Definition solve_pfp (ab: adom t B) (P: PTree.t instruction)
  (entry: node) (transfer: node->instruction->list(node*(t->t)))
  (init: t) : node -> t :=
  let fxp := get_extern_fixpoint entry ab P transfer init in
  if check_fxp entry ab P transfer init fxp then fxp else top.
\end{lstlisting}

The verification of the fixpoint checker yields the following property: the
concretization of the result of the \lstinline,solve_pfp, function is a post-fixpoint of the
concrete transfer function.
%
%\vlnote{Formulation maladroite}
%
That is, given the analysis result~\lstinline,fxp,, for each
node~\lstinline,pc, of the program, applying the corresponding
transfer function \lstinline,tf, to the analysis result yields an abstract value
included in the analysis result.
 
%\vlnote{analyze-entrypoint n’est pas dans le développement}
%%Lemma analyze_entrypoint: forall ab entry f transfer init,
%%  γ init ⊆ γ (analyse ab entry f transfer init entry).
%%Proof. (* proof commands are omitted here *) Qed.
%%
\begin{lstlisting}
Lemma solve_pfp_postfixpoint: forall ab entry P transfer init fxp,
  fxp = solve_pfp ab P entry transfer init →
  ∀ pc i, P[pc] = i →
  ∀ (pc',tf) $\in_{list}$ (transfer pc i), γ(tf(fxp pc))⊆γ(fxp pc').
Proof. (* proof commands are omitted here *) Qed.
\end{lstlisting}

\section{Numerical Abstraction}\label{sec:num-abstract}

Following the design of the \astree{} analyzer~\cite{astreeESOP}, our value analysis is parameterized by a
numerical abstract domain that is unaware of the C memory model.
%\dpnote{Ajouter un peu de texte pour introduire les sous-parties}
We first present the interface of abstract numerical environments,
then how we abstract numerical values in order to build non relational
abstract environments. Finally, we show concrete instances of
numerical domains and how they can be combined. 

\subsection{Abstraction of Numerical Environments}

The first interface captures the notion of numerical environment abstraction.
Given a type \lstinline!t! for abstract values and a notion of variable \lstinline!var! (simple positive integers in our development),
we require an abstract domain that concretizes to \lstinline!℘(var -> int)! and provide
three sound operators \lstinline!range!, \lstinline!assign! and \lstinline!assume!.
\begin{lstlisting}
sign_flag ::= Signed | Unsigned
Definition ints_in_range (r:sign_flag -> itv+⊥) : int := 
                           (γ (r Signed)) ∩ (γ (r Unsigned)).
Record int_dom (t:Type) := {
  int_adom: adom t (var -> int); (* abstract domain structure *)
  (* signed/unsigned range of an expression *)
  range: nexpr -> t -> sign_flag -> itv+⊥;
  range_sound: forall e ρ ab,
    ρ ∈ γ ab -> eval_nexpr ρ e ⊆ ints_in_range (range e ab);
  (* assignment of a variable by a numerical expression *)
  assign: var -> nexpr -> t -> t;
  assign_sound: forall x e ρ n ab,
    ρ ∈ γ ab -> n ∈ eval_nexpr ρ e -> (upd ρ x n) ∈ γ (assign x e ab);
  (* assume a numerical expression evaluates to true *)
  assume: nexpr -> t -> t;  
  assume_sound: forall e ρ ab,
    ρ ∈ γ ab -> Ntrue ∈ eval_nexpr ρ e -> ρ ∈ γ (assume e ab)
}.  
\end{lstlisting}

This interface matches with any implementation of a relational
abstract domain~\cite{CousotHalbwachs78-POPL} on machine %,Mine:thesis
integers. To increase precision, it relies on a notion of expression
tree (type \lstinline!nexpr!) defined as follows and relying on CFG
%unary and binary 
operators.
%\dpnote{tentative de cacher \lstinline!Inductive!}
%described \figref{nexpr}.
%%\begin{lstlisting}
%%nexpr (V:Type) ::=
%%  | NEvar (v:V)
%%  | NEconst (cst:nconstant)
%%  | NEunop (op:unary_operation) (e:nexpr V)
%%  | NEbinop (op:binary_operation) (e1:nexpr V) (e2:nexpr V)
%%  | NEcond (b:nexpr V) (e1:nexpr V) (e2:nexpr V)
%%\end{lstlisting}
%
%\begin{figure}
\begin{equation*}
e_{tr} ::=
    \text{\lstinline,NEvar,}\ \id
\alt \text{\lstinline,NEconst,}\ c
\alt \text{\lstinline,NEunop,}\ \op_1\ e_{tr}
\alt \text{\lstinline,NEbop,}\ \op_2\ e_{tr}\ e_{tr}
\alt \text{\lstinline,NEcond,}\ e_{tr}\ e_{tr}\ e_{tr}
\end{equation*}
  %\caption{\label{fig:nexpr}Numerical Expressions% (polymorphic over the type \lstinline,V, of variables)
%}
%\end{figure}
%
These expressions are associated with a big-step operational semantics
\lstinline!eval_nexpr!
of type 
\lstinline!(var->int) -> nexpr -> ℘(int)! that we define as a partial
function represented by a relation. The semantics is not detailed in
this paper.

\subsection{Building Non-relational Abstraction of Numerical Environments}

Implementing a fully verified relational abstract domain is a
challenge in itself and it is not in the scope of this
paper. We implement instead the previous interface with a standard non
relational abstract environment of the form \lstinline!var -> V♯! where
\lstinline!V♯! abstracts numerical values.
The notion of abstraction of numerical values is captured by the following interface.
\begin{lstlisting}
Record num_dom (t:Type) := { 
  num_adom : adom t int;               (* abstract domain structure *)
  meet: t → t → t+⊥;        (* over-approximation of the concrete *)
  meet_sound: ∀ x y, (γ x) ∩ (γ y) ⊆ γ (meet x y); (* intersection *)
  range: t → sign_flag → itv+⊥;          (* signed/unsigned range *)
  range_sound: ∀ x:t, γ x ⊆ ints_in_range (range x);
  const: constant → t; const_sound:  (*omitted*);
  forward_unop: unary_operation -> t → t+⊥;
  forward_unop_sound: ∀ op x, 
    Eval_unop op (γ x) ⊆ γ (forward_unop op x);
  forward_binop: (* omitted *); forward_binop_sound: (* omitted *);
  backward_unop: (* omitted *); backward_unop_sound: (* omitted *);
  backward_binop: binary_operation -> t → t → t → t+⊥ * t+⊥;
  backward_binop_sound: ∀ op x y z i j k, 
    eval_binop op i j k -> i ∈ (γ x) -> j ∈ (γ y) -> k ∈ (γ z) ->
    let (x',y') := backward_binop op x y z in
      i ∈ (γ x') /\ j ∈ (γ y')
}.
\end{lstlisting}
%binary_operation -> t -> t → t+⊥;
%∀ op x y,    Eval_binop op (γ x) (γ y) ⊆ γ (forward_binop op x y);
%unary_operation -> t → t → t+⊥ ;
It is defined as a carrier~\lstinline!t!, an abstract domain
structure~\lstinline!num_adom! and a bunch of \emph{abstract
  transformers}. Some operators are forward ones: they provide 
properties about the output of an operation. For instance, the
operator~\lstinline!const! builds an abstraction of a single
value. Some operators are backward ones: given some properties about
the input and expected output of an operation, they provide a refined
property about its input.  
Each operator comes with a soundness proof.

We also implement a functor that lifts any abstraction of numerical values into a
numerical environment abstraction. It relies on the functor for finite
reduced maps that we have presented at the end of
Section~\ref{sec:abdom}. Here, \lstinline!PTree! provides an implementation of the
\lstinline!TREE! interface for the \lstinline!var! type. 
\begin{lstlisting}
      NonRelDom.make(t): num_dom t -> int_dom ((PTree.t t)+⊥) 
\end{lstlisting}

The most advanced
operator in this functor is the \lstinline!assume! function. It relies on a backward
abstract semantics of expressions.
\begin{lstlisting}
Fixpoint backward_expr (e:nexpr) (ab:t) (itv:Val) : t := 
  match e with
    | ...
    | NEcond b l r =>
        join
          (backward_expr b (backward_expr r ab itv) (const Nfalse))
          (backward_expr b (backward_expr l ab itv)
               (backward_unop boolval (eval_expr b ab) (const Ntrue)))                
  end.
\end{lstlisting}
We just show and comment the case of conditional expressions.
Given such an expression \lstinline,NEcond b l r,, an abstract
environment~\lstinline,ab, and the expected value~\lstinline,itv, of
this expression, we explore the two branches of the condition. In one
case, the condition~\lstinline,b, evaluated to \lstinline,Nfalse, and
the \emph{right} branch~\lstinline,r, evaluated to~\lstinline,itv,. In
the other case, the condition~\lstinline,b, evaluated to anything
whose boolean value is \lstinline,Ntrue, and the \emph{left}
branch~\lstinline,l, evaluated to \lstinline,itv,. Then we have to
consider that any of the two branches might have been taken, hence the
join. 

% Let’s consider the following example C condition: \lstinline,(x < y ? y < 10 : 0), (such a condition comes from a boolean conjunction of the two comparisons).
% Suppose that at some point in the analysis it is known that this expression evaluates to true (\ie some non zero value), because a branch, guarded by this expression, has just been taken.
% \vlnote{Ça me paraît imbitable.}
% Then the knowledge about the variables \lstinline,x, and~\lstinline,y, can be refined, evaluating this expression backward. On one hand, the backward evaluation of the right branch yields \lstinline,Bot,: no environment can make zero evaluate to true. On the other hand, the backward evaluation of the left branch yield an abstract environment in which the variable \lstinline,y, is known to be not greater than ten; then in this new abstract environment, the expression \lstinline,x < y, is evaluated backward, yielding an abstract environment in which the variable \lstinline,x, is known to be not greater than the variable \lstinline,y, (hence than ten) and the variable \lstinline,y, is known to be not less than the variable \lstinline,x,.

Equipped with such backward operators, the analysis is then able to deal with complex conditions like the following:
%\begin{center}
\lstinline,if (0 <= x && x < y && y < z && z < t, \lstinline,&& t < u && u < v && v < 10),.
%\end{center}
When analysing the true branch of this \lstinline,if,, it is sound to
assume that the condition holds. The backward operator will propagate
this information and infer one bound for each variable. Since backward
evaluation of conditions goes right to left, the following bounds are
inferred: 
\(v<10\), \(u<9\), \(t<8\), \(z<7\), \(y<6\), and \(0\le x<5\). 
Unfortunately, no information is propagated from left to
right. However applying again the \lstinline,assume, function does
propagate information between the various conditions. Iterating this
process finally yields the most precise intervals for all variables
involved in this condition. 

Notice that inferring such precise information is possible thanks to
the availability of complex expressions in the analyzed CFG
program. Compare for example with Frama-C which, prior to any
analysis, destructs boolean operations into nested \lstinline,if,s; it
is thus unable to give both bounds for each variable. 

\subsection{Abstraction of Numerical Values: Instances and Functor}

We gave two instances of the numerical value abstraction interface:
the intervals of signed integers and the intervals of unsigned
integers. 
Several operations are defined on intervals together with their proofs of correctness.
We have to take into account machine arithmetic. We do not try to precisely track integers
that wrap-around intentionally. Instead we systematically test if an overflow may occur and 
fall back to \lstinline!top! when we can't prove the absence of overflow. 
\begin{lstlisting}
Definition repr  (i: itv): itv := if leb i top then i else top.
Definition add (i j: itv): itv :=
  repr { min := i.min + j.min; max := i.max + j.max}.
\end{lstlisting}

We also rely on a reduction operator when the result of an operation may lead to
an empty interval. Since our representation of intervals contains several elements
with the same (empty) concretization, it is important to always use a same representative
for them.\footnote{Otherwise the analyzer may encounter two equivalent values without
noticing it and lose precision.}
\begin{lstlisting}
Definition reduce (min max:Z): itv+⊥ :=
   if min $\leq$ max then NotBot (ITV min max) else Bot.

Definition backward_lt (i j: itv): itv+⊥ * itv+⊥ :=
      (meet i (reduce min_signed (j.max-1)),
       meet j (reduce (i.min+1) max_signed)).
\end{lstlisting}

At run-time, there are no \emph{signed} or \emph{unsigned} integers;
there are only \emph{machine} integers that are bit arrays whose
interpretation may vary depending on the operations they undergo.
Therefore choosing one of the two interval domains may hamper the
precision of the analysis. Consider the following example C program.

\vspace{-9pt}
\begin{center}
\begin{minipage}{.7\textwidth}
%% ATTENTION aux numéros de ligne codés en dur dans
%%  le paragraphe ci-dessous
\begin{lstlisting}[language=C,numbers=left]
int main(void) { signed s; unsigned u;
  if (*) u = $2^{31}$ - 1; else u = $2^{31}$;
  if (*) s = 0; else s = -1;
  return u + s;  }
\end{lstlisting}
\end{minipage}
\end{center}
\vspace{-6pt}
At the end of line~2, an unsigned interval can exactly represent the
two values that the variable~\lstinline,u, may hold. However, the
least signed interval that contains them both is \lstinline,top,. 
Similarly, at the end of line~3, a signed interval can precisely
approximate the content of variable~\lstinline,s, whereas an unsigned
interval would be extremely imprecise. 
Moreover, comparison operations can be precisely translated into
operations over intervals (\eg intersections) only when they share the
same signedness. 
Therefore, so as to get as precise information as possible, we need to
combine the two interval domains. This is done through reduction. 

To combine abstractions of numerical values in a generic and precise way, we
implement a functor that takes two abstractions and a sound reduction operator
and returns a new abstraction based on their reduced product.
\begin{lstlisting}
Definition reduced_product (t t':Type) (N:num_dom t) (N':num_dom t') 
  (R:reduction N N') : num_dom (t*t') := (* omitted definition *)
\end{lstlisting}
A reduction is made of an operator \lstinline!ρ! and a proof that this operator
is a sound reduction.%\dpnote{Je crois que ces notations type-check mais je suis pas sur}
\begin{lstlisting}
Record reduction (A B:Type) (N1:num_dom A) (N2:num_dom B) := { 
  ρ: A+⊥ → B+⊥ → (A * B)+⊥;
  ρ_sound: ∀ a b, (γ a) ∩ (γ b) ⊆ γ (ρ a b)                 }
\end{lstlisting}
Each operator of this functor is implemented by
first using the operator of both 
input domains and then reducing the result with $\rho$.
We hence ensure that each encountered value is systematically 
of the form \lstinline,ρ a b, but we do not prove this fact formally, avoiding 
the heavy manipulation of quotients. Note also that, for soundness purposes, we do
not need to prove that reduction actually reduces (\ie returns a lower element
in the abstract lattice)!

\section{Memory Abstraction}\label{sec:mem-abstract}

The last layer of our modular architecture connects the 
CFG abstract interpreter with numerical abstract domains. It aims at 
translating every C feature into useful information
in the numerical world. On the interpreter side, the interface with
this \emph{abstract memory model} is called \lstinline!mem_dom!. It
consists in trees made of CFG expressions and four basic commands
\lstinline!forget!, \lstinline!assign!, \lstinline!store! and
\lstinline!assume!.

\begin{lstlisting}
Record mem_dom (t:Type) := { (* abstract domain with concretization 
   to local environment and global memory *)
  mem_adom: adom t (env * mem);
  (* consult the range of a local variable *)
  range: t -> ident -> sign_flag -> itv+⊥; 
  range_sound: ∀ ab e m x i,
    (e,m) ∈ γ ab -> (e[x] = vint(i) \/ exists b, e[x] = vptr(b,i)) -> 
    i ∈ (ints_in_range (range ab x));
  (* project the value of a local variable *)
  forget: ident -> t -> t;  
  forget_sound: ∀ x ab, Forget x (γ ab) ⊆ γ (forget x ab);
  (* assign a local variable *)
  assign: ident -> expr -> t -> t;
  assign_sound: ∀ x e ab, Assign x e (γ ab) ⊆ γ (assign x e ab);
  (* assign a memory cell *)
  store: memory_chunk -> expr -> expr -> t -> t;
  store_sound: ∀ κ l r ab, 
    Store κ l r (γ ab) ⊆ γ (store κ l r ab);
  (* assume an expression evaluates to non-zero value *)
  assume: expr  -> t -> t;
  assume_sound: ∀ e ab, Assume e (γ ab) ⊆ γ (assume e ab)
}.  
\end{lstlisting}
Our final analyzer is parameterized by a structure of this type. 
\vspace{-2pt}
%\sbnote{Hum. Comprend-t-on que c'est la fonction du theorem 1 ?}
%\dpnote{Le message important c'est la parametrisation. On n'est pas oblige de donner le type.}
\begin{lstlisting}
value_analysis (t:Type) :  mem_dom t -> 
   program -> node -> (ident -> sign_flag -> Interval.itv +⊥)
\end{lstlisting}
%We use it to
%build a intra-procedural equation system for each function of a program. Each
%equation system is solved by our external fixpoint solver and the result is checked
%by our verified fixpoint checker.\dpnote{Déja dit peut-etre ?} 
A structure of type \lstinline!mem_dom! is built with a functor of the following form.
\vspace{-2pt}
\begin{lstlisting}
  AbMem.make (t:Type) : int_dom t -> mem_dom (t*type_info)
\end{lstlisting}
The numerical abstraction is associated with a flow sensitive type information (of
type \lstinline!type_info!) that we compute at the same time. This type information
tries to recover some information to disambiguate integer and pointer values. 
The abstract domain is built using the product functor presented in Section~\ref{sec:abdom}.
The concretization function of the numeric domain is lifted from a
concretization of type \lstinline!t -> ℘(var->int)!
to a concretization of type \lstinline!t -> ℘(env * mem)! with the
following definition.
\footnote{The types \lstinline{env} and \lstinline{mem} are introduced in Section~\ref{sec:background}.}
%\dpnote{je pense que \lstinline!env! et \lstinline!mem! n'ont pas ete introduits}
\begin{lstlisting}
Definition gamma_mem (ab:t) := λ (e,m):(env*mem). 
    exists ρ:var -> int, ρ ∈ (γ ab) /\
      (∀ x i, (e[x] = vint(i) \/ exists b, e[x] = vptr(b,i)) -> ρ x = i).
\end{lstlisting}

For each transfer function that takes as argument a C expression, we convert it into
a numerical expression in order to feed the numerical abstract
domain. For instance,
the \lstinline!assign! operator takes the following form.
\begin{lstlisting}
Definition assign (id:ident) (e:expr) (ab:t*type_info): t*type_info :=
  let (nm,tp) := ab in  
  (* convert expression e into a numeric form using type infos *)
  match convert tp e with 
    | None => forget id ab (* if we fail, we just project  *)
    | Some ne => 
        (* otherwise we call the numerical assignment operator *)
        (num.assign id ne nm, ... (* type info update omitted *))
end.  
\end{lstlisting}

Removing some ambiguity between pointers and integers is mandatory for
soundness. As an example, consider the unsigned equality expression
\lstinline!(x ==u y)!. For the sake of precision of the analysis, it
is important to convert it into a simple numerical
equality \lstinline!x == y! before using the \lstinline!assume!
operator of the numerical abstract domain. However if \lstinline!x!
contains a numerical value and \lstinline!y! a pointer, the first
formula is always false while assuming the second formula in the
numerical world would lead to a spurious assumption about the offset
of the pointer in \lstinline!y!.

%\input{valanal} %old version

%\section{Soundness proof}\label{sec:proof}
%\sbnote{Enlever cette section 6}
%\input{proof}

\section{Experimental Evaluation}\label{sec:exper-eval}

% We have integrated our value analysis in the CompCert 1.11 compiler. Our analysis
% computes numerical invariants (?) in programs expressed in the CFG internal 
% representation of the CompCert compiler. The value analysis takes as input a CFG 
% program (obtained from a C file by the CompCert compiler) and outputs for each 
% program point an invariant over local variables in this program 
% point.

Our verified value analyzer takes as input a CFG program and outputs ranges
for every variable at every point of the program. 
Our formal development adds about 7,500 lines of Coq code (consisting of
4,000 lines of Coq functions and definitions and 3,500 lines of Coq statements 
and proof scripts) and 200 lines of OCaml to the 100,000 lines of Coq and 1,000 
lines of OCaml provided in CompCert 1.11.

We have conducted some experiments to evaluate the precision and the
efficiency of our analyzer. 
%We perform an experimental evaluation for two reasons.
%
%First, we want 
%to ensure that the precision of the analyzer is adequate; 
Indeed, an analyzer that always returns ``top'' is easily
proved correct, but useless. It is therefore important to distinguish
between bounded and unbounded variables.
Moreover, a precise but non-scalable analyzer has
limited applicability.
In order to evaluate the precision and efficiency of our value
analysis, we use the OCaml extracted code to
compile our benchmark programs into CFG programs
%After adding an inlining pass to perform 
%the equivalent of an interprocedural analysis, 
and to run our analyzer on them.

We compare our analyzer to two interval-based analyzers operating over C programs: a 
state-of-the-art industrial tool, Frama-C~\cite{framac}, and an implementation 
of a value-range analyzer~\cite{aplas12}. % (which will be referred to as {\em Wrapped}).
Frama-C is an industrial-strength framework for static analysis, 
developed at CEA. It integrates an abstract interpretation-based interprocedural
value analysis on interval domains with congruence, k-sets and memory analysis.
It operates over C programs and has a very deep knowledge
of its semantics, allowing it to slice out undefined behaviors for more precise
results. It currently does not handle recursive functions.  
The value-range analyzer, which will be referred to as Wrapped is described in~\cite{aplas12}.
It relies on LLVM and operates over its intermediate representation 
to perform an interval analysis in a signedness-agnostic manner,
using so-called ``wrapped'' intervals to deal with machine integer issues such as 
overflows while retaining precision. It is an intraprocedural tool, but can benefit
from LLVM's inlining to perform interprocedurally in the absence of recursion.

The 3 tools have been compared on significant C programs from CompCert's test suite.
They range from a few dozen to a few thousand statements. %In
%order to compare sound analyses, we commented out recursive calls in a
%few programs, since Frama-C's value analysis currently does not handle them. 
%\andnote{Justifier pourquoi on n' a pas pris tous les programmes ? Floats, pointeurs, etc.}
%
%It is not straightforward to compare analyses which operate over close
%but different languages and do not rely on the same semantics. 
To
relate information from different analyses, we annotated the 
programs to capture information on integer variables at function
entries and exits and at loops (for iteration variables). This amounts
to 545 annotations in the 20 programs considered. For each program
point, we counted the number of bounded variables.
%
%The definition of a ``bounded'' variable must take into account some aspects of 
%the analysis. 
We consider as bounded any variable 
whose inferred interval has no more than \(2^{31}\) elements,
and hence rule out useless intervals like \lstinline!x∈[$-2^{31}$,$2^{31}-2$]!,
inferred after a guard like \lstinline!x<y!. 
%
% For instance, an inequality $x < y$ between variables containing 
% unknown values (e.g., $x,y \in [-2^{31},2^{31}-1]$, if both are unsigned) 
% allow the analyzer to infer that, after the comparison, one of those $2^{32}$ values is not 
% allowed for each of the variables in question ($y > -2^{31}$ and $x < 2^{31}-1$). 
% Such an imprecise information is in itself mostly useless, and so such variables 
% were not considered bounded. Similarly, if there are more comparisons, further 
% values may be removed from each variable domain, making it hard to establish a 
% precise interval. Having in mind the fact that even large intervals such as 
% $[0,2^{31}-1]$ and $[-2^{31},2^{31}-2] \mod 2$ (that is, all even integers) may be 
% useful, 
%
%  We inspected the results to verify that it is indeed never the case 
% that a larger interval gives useful information to the analysis 
% (for instance, there are no cases where $[-1,2^{31}-1]$ is found).
%
Finally, to be able to compare the results of an interprocedural analysis with
those of two intraprocedural analyses with inlining, we considered for each
annotation the union of the intervals of all call contexts. Less than 10\% of intervals
present a union of different intervals, and among those several preserve the boundedness
for all contexts. Overall, its impact on the results is negligible.

\begin{figure}%[!ht]
  \centering
\includegraphics[width=.95\textwidth]{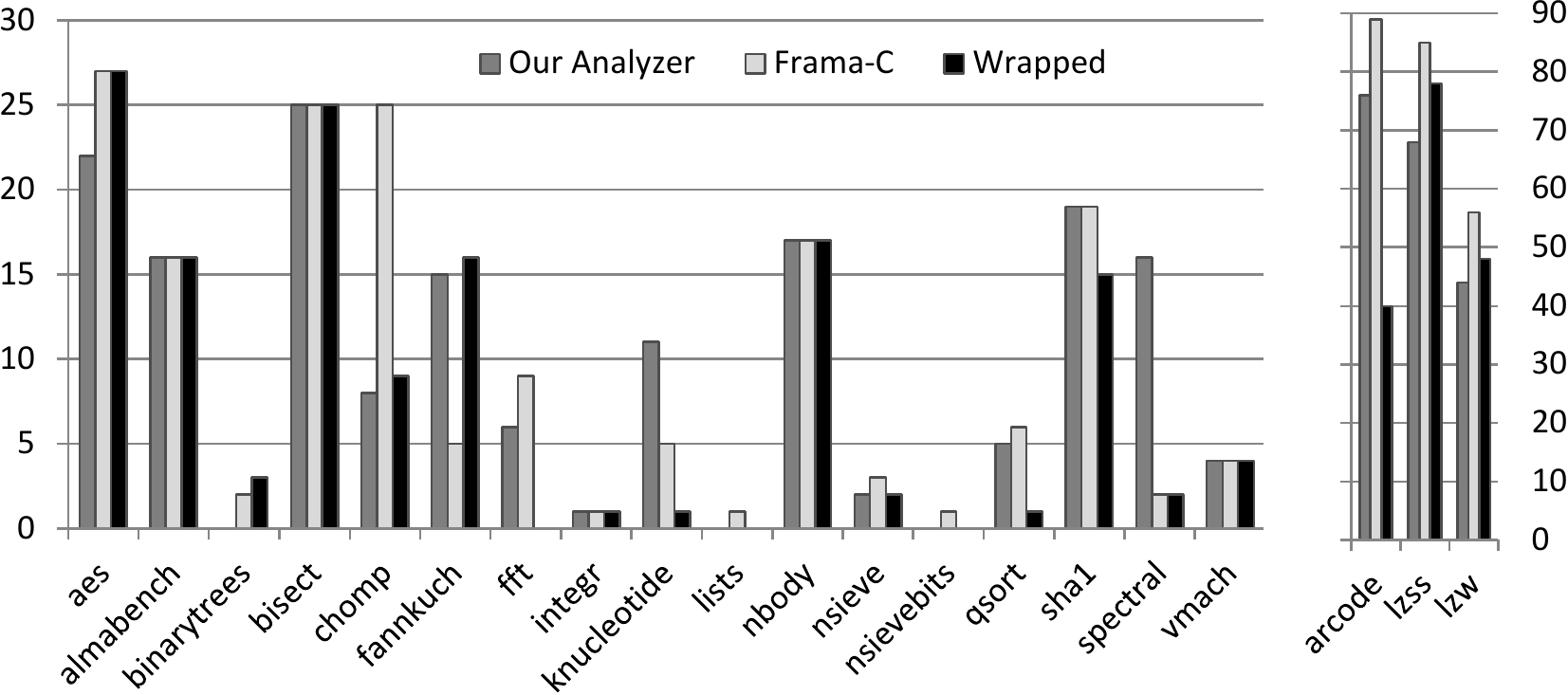}
\caption{Number of bounded intervals (bounded per program and analyzer).}
\label{fig-results}
\end{figure}

The results are shown in Figure~\ref{fig-results}, 
which displays the number of bounded variables per program
and per analyzer. In total, Frama-C bounded 398 variables, our analyzer got 355, 
and Wrapped ended up with 305.
The main differences between our analyzer and Frama-C, especially on the larger
benchmarks (\lstinline,lzw,, \lstinline,arcode, and \lstinline,lzss,)
result from global variable tracking and congruence information. Such reasoning
is not handled by our analyzer. On the other hand, the precision of our product
of signed and unsigned domains allows us to bound more variables (\eg on 
\lstinline,fannkuch,), where Wrapped also obtains a good score, mainly due to
variables bounded as $[0,2^{31}-1]$ and similar values. Some issues
with the inlining used by Wrapped explain its worse results in \lstinline,fft,,
\lstinline,knucleotide, and \lstinline,spectral,.
% /andnote{Suggestion à rajouter pour élever notre morale}
% Without the last 3 programs, 
% scores are more evenly distributed: 154 for Frama-C, 150 for
% our analyzer and 125 for Wrapped.

%We could not identify the cause of the precision loss for Wrapped on fft, 
%knucleotide and spectral.\dpnote{phrase trop negative}

We also compared the execution times of the analyses. 
%, mainly to determine if ours is efficient enough. 
Overall, our analysis runs faster than Frama-C
because we track less information, such as pointers and global variables.
For programs without these features, both analyses run in roughly the same time,
from a few tenths of seconds for the smaller programs up to a few seconds for the larger
ones. Wrapped's analysis is faster than the others. 
On a larger benchmark (over 3,000 lines of C code and 
about 10,000 CFG instructions after inlining) our analysis took 34 seconds to
perform.

It is hard to draw final conclusions about the precision of our tool
from these experiments. Frama-C, for instance, is likely to perform better on
specific industrial critical software for which it has been specially tuned. 
Nevertheless we give evidence that our analyzer performs non-trivial reasoning
over the C semantics, close to that of state-of-the-art (non-verified) tools.

% Overall, we managed to confirm our extracted code is sufficiently precise and
% efficient in practice. The comparison also allowed us to quantify the gain for
% each new feature that will included in future versions, such as congruence 
% information and global variable tracking.

\section{Related Work}\label{sec:related-work}

While mechanization of research paper proofs attracts an increasing
number of practitioners, %~\cite{AydemirBFFPSVWWZ05}, 
it should not be
confused with the activity of developing a formally verified compiler
or static analyzer.  
% In this area, the CompCert compiler~\cite{Leroy-backend} is a
% landmark: it is a fully verified moderately optimizing compiler for a
% realistic subset of C, and the extracted code of this compiler is
% quite competitive with industrial non-verified compilers that are used for compiling
% safety-critical software. Our work is initially
% inspired by this achievement and our target is
% the area of abstract-interpretation-based static analyzers.
Our work is initially
inspired by the achievement of the CompCert compiler~\cite{Leroy-backend}
but we target 
the area of abstract-interpretation-based static analyzers.

Previous work on mechanized verification of static analyses has been
mostly based on classic data flow frameworks: Klein and Nipkow
instantiate this framework for inference of Java bytecode
types~\cite{KleinN-TOPLAS}; Coupet-Grimal and
Delobel~\cite{Coupet-Grimal-Delobel-05} and
Bertot\etal~\cite{Bertot-Gregoire-Leroy-05} for compiler
optimizations, and Cachera\etal~\cite{Cachera-Jensen-05} for control
flow analysis. Vafeiadis\etal~\cite{Vafeiadis:2011} rely on a simple
data flow analysis to verify a fence elimination optimization for
concurrent C programs.  
%Their contribution is more focused on the
%challenging semantic proof they have to perform than on the
%algorithms of their analyzer.
%
Compared to these prior works, our value analysis relies 
on fixpoint iterations that are accelerating with widening
operators. Cachera and Pichardie~\cite{ITP10:Cachera:Pichardie} and
Nipkow~\cite{DBLP:conf/itp/Nipkow12} describe a verified static
analysis based on widenings but their technique is restricted to
structured programs and targets languages without machine arithmetic
nor pointers.  
%
%Our loop bound estimation would gain extra precision by relying
%on alias analysis. In this area 
Leroy and Robert~\cite{Leroy:CPP12} have developed a
points-to analysis in the CompCert framework. This static analysis 
technique is quite orthogonal to what we formalize here. Their verified tool
is not compared to any existing analyzer. 
%Such alias information
%could be reused for a more fine-grain slicing of memory operations.
%
Hofmann\etal~\cite{Hofmann:2010} provide a machine-checked
correctness proof in Coq for a generic post-fixpoint solver named RLD.  
%The
%solver takes as input a set of constraints and finds a suitable
%solution by implementing an optimized dynamic iteration strategy.
The formalized algorithm is not fully executable and cannot be extracted 
to OCaml code.

%Our value analysis provides a well specified entry for a relational numerical
%abstract domain but we do not try to formalize such a domain in this paper.
%Besson\etal~\cite{TGC10:Besson:Jensen:Pichardie:Turpin} provide a technique
%to verify the result of a polyhedral abstract domain rather than verifying
%its implementation. 

Of course the area of non-verified static analysis for C programs is
a broader topic.  In our context, the most relevant and inspiring works
are the static analyses devoted to a precise handling of signed and
unsigned integers~\cite{simon07taming,aplas12} and %papers on
 the \astree{} static analyzer~\cite{astreeESOP}. %,mine-FMSD09,airbus2,Cousot06ASIAN}.
Our current formalization is directly inspired by \astree's design choices, trying
to capture some of its key interfaces. Our current abstract memory model is aligned
with the model developed by Min\'e~\cite{mine-LCTES06} because we connect
a C abstract semantics with a generic notion of numerical abstract domain. 
Still our treatment of memory is simplified since we only track values of local variables
in the current implementation of our analyzer.

% Our work is significant for many reasons. 
% It constitutes the first machine-checked proof of a nontrivial
% value analysis based on abstract interpretation and a reference implementation of a
% tool. 
% In addition, formalizing such a tool requires hard work on the soundness proof. 
% Furthermore, the algorithm we prove is working on unstructured control
% flow graphs. These graphs have specific properties that must be kept in
% mind along the specification of the algorithm. Finally, we took a special care of the
% algorithmic complexity of the generated code since it deals with a real and
% concrete problem, value analysis that has been integrated to the CompCert
% compiler. 

\section{Conclusion}\label{sec:conclusion}

This work provides the first verified value analysis for a realistic
language as C.  Implementing a precise value analysis for C is highly
error-prone. We hope that our work shows the feasibility of developing
such a tool together with a machine-checked proof. 
The precision of the analysis has been experimentally evaluated and
compared on several benchmarks.  The paper’s technology performs
comparably to existing off-the-shelf (unverified!) tools, Frama-C~\cite{framac}
and Wrapped~\cite{aplas12}.
Our contribution is also methodological. 
Our formalization, its lightweight interfaces and its proofs can be easily reused
to develop different formally verified analyses.
%Formalizing such a tool can
%be considerably speed-up when adopting the various weak specifications
%we have shown in this paper.

Now that the main interfaces are defined, we expect to improve our
analyzer in several challenging directions.
First, we want to
replace the current memory abstraction with a domain similar
to Miné’s memory model~\cite{mine-LCTES06}. Verifying such a domain
raises specific challenges not only in terms of semantic proofs but also
in terms of efficient implementation of the transfer
functions. Without special care,
the domain may not be able to scale to large enough programs. 
We also intend to connect relational abstract domains to the interface
for numerical environments. We would like to develop efficient
validation techniques following
Besson\etal~\cite{TGC10:Besson:Jensen:Pichardie:Turpin} approach and
test their 
efficiency on large programs. 
The last and important challenge concerns floats. \astree{} relies on
subtle reasoning and manipulation on floats. CompCert has recently been
enhanced with a fully verified implementation of floating-point
arithmetic~\cite{Boldo-Jourdan-Leroy-Melquiond-2013} and we hope to be able to 
incorporate them in our own value analysis.

\subsection*{Acknowledgements}
We thank Antoine Min\'e, David Monniaux and Xavier Rival for many fruitful discussions
on the Astr\'ee static analyzer. We thank Jacques-Henri Jourdan and
Xavier Leroy for integrating the CFG language into the CompCert compiler.

% \bibliographystyle{plain}
% \bibliography{biblio}

%\begin{thebibliography}{5}

%\end{thebibliography}

%\section*{Appendix}
%\input{appendix}

\end{document}